\documentclass[12pt,a4paper]{article}
\usepackage{amsmath}
\usepackage{amssymb} 
\usepackage{graphicx} 
\usepackage{float}
\restylefloat{table}
\begin{document}
\begin{titlepage} 
\begin{flushright} IFUP--TH/2018\\ 
\end{flushright} ~
\vskip .8truecm 
\begin{center} 
\Large\bf Torus classical conformal blocks
\end{center}
\vskip 1.2truecm 
\begin{center}
{Pietro Menotti} \\ 
{\small\it Dipartimento di Fisica, Universit{\`a} di Pisa}\\ 
{\small\it 
Largo B. Pontecorvo 3, I-56127, Pisa, Italy}\\
{\small\it e-mail: pietro.menotti@unipi.it}\\ 
\end{center} 
\vskip 0.8truecm
\centerline{ May 2018}
                
\vskip 1.2truecm
                                                              
\begin{abstract}
After deriving the classical Ward identity for the variation of the
action under a change of the modulus of the torus we map the problem
of the sphere with four sources to the torus.  We extend the method
previously developed for computing the classical conformal blocks for 
the sphere topology to the tours topology. We
give the explicit results for the classical blocks up to the third
order in the nome included and compare them with the classical limit of
the quantum conformal blocks. The extension to higher orders is
straightforward.

\end{abstract}

\end{titlepage}
 
\eject

\section{Introduction}

After the seminal papers of Zamolodchikov and Zamolodchikov
\cite{ZZ,AlZ1,AlZ2} a lot of work has been performed about the
structure and computation of the conformal blocks both of the sphere
and higher genus surfaces
\cite{onofri,hadasz,MMM,MMS,piatek,perlmutter,
  beccaria,alkalaevbelavin,PTY}.

In a previous paper \cite{CCblocks} a simple iterative method was
developed to compute the classical conformal block for the sphere
topology. Essentially the idea was to expand the Heun equation in the
parameter $x$ which represents the modulus and to match such an
expansion with the one obtained by performing the classical
transformation of the equation with $x=0$, the hypergeometric
equation, when one operates on it a proper transformation of
the variable.

In this paper we want to show how the same method can be applied to
the one point torus amplitude. The procedure
is simpler due to the special value assumed by three of the four
singularities. To reach order $n$ in the expansion
in the nome $\tilde q=e^{2\pi i\tau}$ one has to work out the
corresponding expansion on the sphere to order $2n$ in $x$.

To relate such conformal block expansion with the quantum problem we
work out the classical Ward identity, which gives the change of the
classical action under a change of the modulus $\tau$,
in terms of the accessory parameter of the problem. The procedure is rigorous
and the equation we obtain agrees with the one derived by Piatek
\cite{piatek} from the classical limit of the linear differential
equation obeyed by a correlation function containing a null field
which is a consequence of the quantum Ward identity proved Eguchi and
Ooguri \cite{eguchiooguri}.

Then we proceed to the comparison with the classical limit of the
known quantum conformal blocks finding full agreement.

\section{The classical Ward identity}

In this section we give a direct proof of the Ward identity relating
the change of the torus action under a change of the modulus in terms 
of the accessory parameter and the Weierstrass $\eta_1$ function.

The Liouville action on the torus represented by a parallelogram in
the $z$-plane with periodic boundary conditions, is given by
\begin{equation}\label{SzactionTorus}
S_z=\frac{1}{2\pi}\int_{T}(\frac{1}{2}
\partial\phi\wedge\bar \partial\phi+e^\phi dz\wedge d\bar z)\frac{i}{2}-
\frac{\eta_K}{4\pi i}\oint_{\epsilon_K}\phi(\frac{dz}{z-z_K}-
\frac{d\bar z}{\bar z-\bar z_K})-\eta_K^2\log\epsilon^2_K
\end{equation}
where $z_K$ are the location of the sources of strength $\eta_K<1/2$
i.e. elliptic singularities.  Extension to parabolic singularities
$\eta_K=\frac{1}{2}$ poses no problem \cite{HG,existence}.  
The normalization of the action $S_z$
is the same as the one adopted in \cite{ZZ} and \cite{HG}.

Working with periodic boundary conditions is rather cumbersome; so we
go over the Weierstrass representation of the torus given by the equations
\begin{equation}\label{wpdefinition}
u=\wp(z),~~ w^2=4(u-e_1)(u-e_2)(u-e_3),~~~~e_j=\wp(\omega_j),~~~~~e_1+e_2+e_3=0~.
\end{equation}
Due to the invariance of
the area i.e. $e^\varphi du\wedge d\bar u=e^\phi dz\wedge d\bar z$,
in the $u$-representation the field is given by
\begin{equation}\label{phivarphitorus}
\varphi(u) =\phi(z)+\log J\bar J,~~~~~~~~J=\frac{dz}{du}~.
\end{equation}
Substituting in eq.(\ref{SzactionTorus}) we obtain \cite{HG} 
\begin{equation}\label{SzSu2}
S_z=S_u-\frac{1}{2}\sum_K\eta_K(1-\eta_K) \log|4
(u_K-e_1)(u_K-e_2)(u_K-e_3)|^2-\frac{1}{4}\log|
(e_1-e_2)(e_2-e_3)(e_3-e_1)|^2
\end{equation}
where
\begin{eqnarray}\label{SuTorus}
&&S_u= 
\frac{1}{2\pi}\int_{D_\varepsilon}(\frac{1}{2}\partial\varphi\wedge\bar
\partial\varphi+ e^\varphi du\wedge d\bar u)
\frac{i}{2}
-\frac{\eta_K}{4\pi i}\oint_{\varepsilon_K} \varphi(\frac{du}{u-u_K} -\frac{d\bar
  u}{\bar u-\bar u_K})-\eta_K^2\log\varepsilon_K^2\nonumber\\
&-&\frac{1}{16\pi i}\oint^d_{\varepsilon_l}\varphi(\frac{du}{u-e_l} 
-\frac{d\bar  u}{\bar u-\bar e_l}) 
-\frac{1}{8}\log\varepsilon_l^2
+\frac{1}{8\pi i}\frac{3}{2}\oint^d_{R}\varphi(\frac{du}{u} 
-\frac{d\bar  u}{\bar u}) 
+\frac{1}{2}\bigg(\frac{3}{2}\bigg)^2\log R^2
\end{eqnarray}
where $D_\varepsilon$ is the double sheeted plane 
and the index $d$ on the contour integrals means that a double turn 
has to be taken around the kinematical
singularities $e_l$, $l=1,2,3$ and at $\infty$ in order to come back
to the starting point.
For a single source of strength $\eta_s$ placed 
at the origin $z=0$ i.e. $u=\infty$ we have \cite{torusIII}
\begin{eqnarray}\label{Suorigin}
&&S_u= 
\frac{1}{2\pi}\int_{D_\varepsilon}(\frac{1}{2}\partial\varphi\wedge\bar
\partial\varphi+ e^\varphi du\wedge d\bar u)
\frac{i}{2}
-\frac{1}{16\pi i}\oint^d_{\varepsilon_l}\varphi(\frac{du}{u-e_l} 
-\frac{d\bar  u}{\bar u-\bar e_l}) 
-\frac{1}{8}\log\varepsilon_l^2\nonumber\\
&-&\frac{1}{8\pi i}(\eta_s -\frac{3}{2})\oint^d_{R}\varphi(\frac{du}{u} 
-\frac{d\bar  u}{\bar u}) 
+\frac{1}{2}\big(\eta_s-\frac{3}{2}\big)^2\log R^2
\end{eqnarray}
and
\begin{equation}\label{SzSu}
S_z = S_u -\frac{1}{4}\log|(e_1-e_2)(e_2-e_3)(e_3-e_1)|^2 -2 \eta_s
\log 2~.
\end{equation}
The Liouville equation for $\phi$ is
\begin{equation}
-\partial_z\partial_{\bar z}\phi+e^\phi = 2\pi\eta_s\delta^2(z),~~~~
0<\eta_s<\frac{1}{2}~.
\end{equation}
The auxiliary equation in the $z$-representation is given by
\begin{equation}
f''(z)+\delta_s\big(\wp(z)+\beta\big)f(z)=0
\end{equation}
where $\wp$ is Weierstrass's $\wp$ function and $\beta$ the accessory
parameter and $\delta_s=\eta_s(1-\eta_s)>0$.
The auxiliary equation in the $u$ variable is given by $f''+Q_uf=0$
where by standard procedure \cite{torusI} we have
\begin{eqnarray}\label{Qu}
& &Q_u(u) = \frac{\delta_s}{4}\frac{u+\beta}
{(u-e_1)(u-e_2)(u-e_3)}+
\frac{3}{16}\bigg(\frac{1}{(u-e_1)^2}+\frac{1}{(u-e_2)^2}+
\frac{1}{(u-e_3)^2}
\\
&- &\frac{2e_1}{(e_1-e_2)(e_1-e_3)(u-e_1)}-
\frac{2e_2}{(e_2-e_1)(e_2-e_3)(u-e_2)}-
\frac{2e_3}{(e_3-e_1)(e_3-e_2)(u-e_3)}\bigg)\nonumber~.
\end{eqnarray}  
We recall that the complex parameters $e_j=\wp(\omega_j)$ are related
by $e_1+e_2+e_3=0$. Both the half-period $\omega_1$ and the modulus
$\tau$ are functions of the $e_j$. At the end we
are interested in the change of the action (\ref{SzactionTorus}) under
a change of the modulus $\tau$ i.e. under a change of the $e_j$
keeping and the half period $\omega_1$ fixed.  
The result for the change of the action under a change of 
the position of the sources is
\begin{equation}\label{polyakov}
\frac{\partial S_u}{\partial u_K} = -\frac{B_K}{2}
\end{equation}
where $B_K$ are the accessory parameters of $Q_u$ at the positions
$u_K$, i.e. $B_K/2$ is the residue of the simple pole of $Q_u$ at 
$u_K$. This is the well known Polyakov relation.
For a change of $e_j$ we have \cite{HG}
\begin{equation}\label{polyakovej}
\frac{\partial S_u}{\partial e_j} = -B_j~.
\end{equation}
Notice the factor $2$ of difference between eq.(\ref{polyakov}) and 
eq.(\ref{polyakovej}). 
For completeness
we give in Appendix A a short derivation of (\ref{polyakovej}) which
holds for general hyperelliptic surfaces.
In particular for the case of the torus with a single source we have
using (\ref{SzSu})
\begin{equation}
\frac{\partial S_z}{\partial e_1}
=-\frac{\delta_s}{2}\frac{\beta+e_1}{(e_1-e_2)(e_1-e_3)}
\end{equation}
and similar results for the derivative w.r.t. $e_2$ and $e_3$. As we
mentioned before we are interested in the variation of $S_z$ when
$\omega_1$ is kept fixed. Under this conditions the $e_j$ become
functions of the modulus $\tau$ and the change of the action $S_z$ is
given by
\begin{equation}\label{deltaSz}
\delta S_z = -\frac{\delta_s}{2}\bigg[
\frac{\beta+e_1}{(e_1-e_2)(e_1-e_3)}\delta e_1
+\frac{\beta+e_2}{(e_2-e_1)(e_2-e_3)}\delta e_2
+\frac{\beta+e_3}{(e_3-e_1)(e_3-e_2)}\delta e_3\bigg]
\end{equation}
under the constraints $\omega_1={\rm const}$ and $\delta e_1+\delta
e_2+\delta e_3=0$.
To compute (\ref{deltaSz}) we exploit the relation \cite{BatemanII}
\footnote{We use here and in the following the definition of
the theta functions $\theta_j$ adopted in \cite{DLMF,Wolfram} and not
the one of \cite{BatemanII}}
\begin{equation}\label{discriminantrelation}
\Delta^{1/4}=\frac{\pi^3}{4\omega^3_1}(\theta_1'(0|\tau))^2
\end{equation}
being $\theta_1$ the elliptic theta function 
and $\Delta$ the discriminant
\begin{equation}
\Delta = 16 (e_1-e_2)^2(e_2-e_3)^2(e_3-e_1)^2~.
\end{equation}
Taking the logarithmic differential of (\ref{discriminantrelation}) 
at constant $\omega_1$ we
have
\begin{eqnarray}\label{logdiff}
&&\frac{3}{2}\bigg(\frac{e_1}{(e_1-e_2)(e_1-e_3)}de_1+
\frac{e_2}{(e_2-e_1)(e_2-e_3)}de_2+
\frac{e_3}{(e_3-e_2)(e_3-e_1)}de_3\bigg)=\nonumber\\
&=&2\frac{\dot\theta'_1(0|\tau)}{\theta'_1(0|\tau)}d\tau
= -i\frac{\pi}{2}\frac{\theta_1'''(0|\tau)}{\theta'_1(0|\tau)}d\tau
=i\frac{6\omega_1}{\pi}\eta_1d\tau
\end{eqnarray}
where the dot denotes the derivative w.r.t. $\tau$ and we used \cite{DLMF}
\begin{equation}\label{heattheta}
\dot\theta_j(z|\tau) = -i\frac{\pi}{4}\theta_j''(z|\tau)
\end{equation}
and \cite{BatemanII}
\begin{equation}
\frac{\theta_1'''(0|\tau)}{\theta_1'(0|\tau)} 
= -\frac{12 \omega_1}{\pi^2}\eta_1
\end{equation}
where $\eta_1=\zeta(\omega_1)$. This gives the sum of the three terms in 
eq.(\ref{deltaSz})
with the $e_j$ at the numerator. For the three terms proportional to
$\beta$ we notice that from 
\begin{equation}
e_1-e_2=\frac{\pi^2}{4\omega_1^2}\theta_4^4(0|\tau),~~~~q=e^{i\pi\tau}
\end{equation}
we have
\begin{equation}\label{de1mde2}
\frac{de_1-de_2}{e_1-e_2}= 4\frac{\dot\theta_4(0|\tau)}{\theta_4(0|\tau)}d\tau=
-i\pi\frac{\theta''_4(0,q)}{\theta_4(0,q)}d\tau
\end{equation}
and similarly
\begin{equation}\label{de2mde3}
\frac{de_2-de_3}{e_2-e_3}= 4\frac{\dot\theta_2(0|\tau)}{\theta_2(0|\tau)}d\tau=
-i\pi\frac{\theta''_2(0,q)}{\theta_2(0,q)}d\tau
\end{equation}
where we used again eq.(\ref{heattheta}). We employ now the formula
\cite{Wolfram}
\begin{equation}
\eta_1 =
-e_i\omega_1-\frac{\pi^2}{4\omega_1}\frac{\theta''_{i+1}(0,q)}
{\theta_{i+1}(0,q)}
\end{equation}
which by subtraction gives
\begin{equation}\label{e1me3}
\omega_1 (e_1-e_3) =
\frac{\pi^2}{4\omega_1}\bigg(\frac{\theta''_4(0,q)}
{\theta_4(0,q)}-\frac{\theta''_2(0,q)}
{\theta_2(0,q)}\bigg)
\end{equation}
and so using eqs.(\ref{de1mde2},\ref{de2mde3}) and eq.(\ref{e1me3}) we have
\begin{eqnarray}
&&\frac{de_1-de_2}{(e_1-e_2)(e_1-e_3)}-
\frac{de_2-de_3}{(e_2-e_3)(e_1-e_3)}\\
&=&
\frac{de_1}{(e_1-e_2)(e_1-e_3)}+\frac{de_2}{(e_2-e_1)(e_2-e_3)} +
\frac{de_3}{(e_3-e_1)(e_3-e_2)} =
-4i \frac{\omega_1^2}{\pi} d\tau~.\nonumber
\end{eqnarray}
Combining with eq.(\ref{logdiff}) 
we obtain
\begin{equation}\label{classWard}
\frac{\partial S_z}{\partial\tau}\bigg|_{\omega_1}=
\frac{
2i \delta_s\omega_1^2}{\pi} (\beta-\frac{\eta_1}{\omega_1})~.
\end{equation}
Such equation was derived by Piatek \cite{piatek} exploiting the
classical limit of a linear differential equation containing the
null-field $\chi_{-\frac{b}{2}}$, consequence of the quantum Ward
identity given by Eguchi and Ooguri \cite{eguchiooguri}.  Here it has
been derived directly from the classical action.

\section{The torus blocks}

In \cite{CCblocks} a simple iterative method was developed to compute
the classical conformal block for the sphere topology with four
sources.  The auxiliary differential equation of the $4$-point
Liouville problem is given by
\begin{equation}\label{auxiliary}
y''(z)+Q(z)y(z)=0
\end{equation}
with
\begin{equation}\label{sphereQ}
Q =\frac{\delta_0}{z^2}+ \frac{\delta}{(z-x)^2}+
\frac{\delta_1}{(z-1)^2}
+\frac{\delta_\infty-\delta_0-\delta-\delta_1}{z(z-1)}-
    \frac{C(x)}{z(z-x)(z-1)}
\end{equation}
where $\delta_j=(1-\lambda_j^2)/4$ and $C(x)$ is the accessory
parameter.

The main idea of \cite{CCblocks,sphere4} 
is to expand the $Q$ appearing in
eq.(\ref{sphereQ}) in powers of $x$ and to match such an expansion
with the one obtained from $Q_0$
\begin{equation}\label{Q0}
Q_0 = \frac{\delta_\nu}{z^2}+\frac{\delta_1}{(z-1)^2}+
    \frac{\delta_\infty-\delta_1-\delta_\nu}{z(z-1)}
\end{equation}
after performing on $Q_0$ the transformation
\begin{equation}\label{transformation}
z(v,x)= \frac{v- {\cal C}-{\cal B}_1/v-{\cal B}_2/v^2+\dots}
{1- {\cal C}-{\cal B}_1-{\cal B}_2+\dots}
\end{equation}
where
\begin{eqnarray}
{\cal C} &=& xc_1+x^2 c_2 + x^3 c_3+\dots\nonumber\\
{\cal B}_1 &=& x^2 b_{11}+x^3 b_{12}+x^4 b_{13}+\dots\nonumber\\
{\cal B}_2 &=& x^3 b_{21}+x^4 b_{22}+x^5 b_{23}+\dots\nonumber\\
{\cal B}_3 &=& x^4 b_{31}+x^5 b_{32}+x^6 b_{33}+\dots\nonumber\\
&&.......................
\end{eqnarray} 
The transformed $Q$ according to the well known rules is
\begin{equation}
Q_v(v) = Q_0(z(v))\big(\frac{dz}{dv}\big)^2-\{z,v\}
\end{equation}
where $\{z,v\}$ is the Schwarz derivative of $z$ w.r.t. $v$. Such a
method generates an iterative structure for the coefficients $c_j$
and $b_{jk}$ and the derivatives of $C$ w.r.t. $x$, $C^{(n)}$ 
\cite{CCblocks}. 
The procedure
apart the known problem of the convergence of the series, is rigorous.
As an example the
explicit form of the expansion of the
accessory parameter $C(x)$ up to the third order included was given 
\cite{CCblocks}.

We want now to adapt such a structure to the computation of the
classical conformal blocks for the torus with one source.

The same $Q$ which describes the sphere can be used to describe the
torus provided we perform the following substitutions
\begin{equation}\label{delta316}
\delta_0=\delta=\delta_1=3/16 
\end{equation}
and $\delta_\infty = \frac{1-\lambda_s^2}{16}+\frac{3}{16}$
where the source dimension $\delta_s$ appearing in the previous section
is given by $\delta_s = (1-\lambda_s^2)/4$. The value of the $\delta$'s
appearing in eq.(\ref{delta316}) induce the two-sheet structure of the
$u$-plane describing the torus.

In order to exploit the results of \cite{CCblocks} we maneuver
eq.(\ref{Qu}) in the form (\ref{sphereQ}) to obtain
\begin{eqnarray}\label{Qv}
&&Q_v(v) =\frac{\delta_s}{4}\frac{v+\frac{\beta+e_2}{e_1-e_2}}{v(v-1)(v-x)}+
\frac{3}{16}\bigg(\frac{1}{v^2}+\frac{1}{(v-1)^2}+\frac{1}{(v-x)^2}\nonumber\\
&+&\frac{2e_1}{(e_1-e_2)(x-1)(v-1)}-\frac{2e_2}{xv(e_1-e_2)}-
\frac{2e_3}{x(x-1)(e_1-e_2)(v-x)}
\bigg)
\end{eqnarray}
where $x$ is related the the $e_j$ by
\begin{equation}
x =\frac{e_3-e_2}{e_1-e_2}~~~~{\rm and}~~v=\frac{u-e_2}{e_1-e_2}~.
\end{equation}
Eq.(\ref{Qv}) compared to eq.(\ref{sphereQ}) gives
\begin{equation}\label{betaC}
\beta = -e_3 +\frac{4}{\delta_s}
\bigg(-C\big(\frac{e_3-e_2}{e_1-e_2},\delta_\nu\big)+
\frac{3}{8}\frac{e_3}{(e_1-e_2)}\bigg)(e_1-e_2)~.
\end{equation}
The procedure originally adopted by \cite{ZZ,AlZ1} is to start from a
generic monodromy along the loop $M$ embracing the origin and the
point $x$ and whose trace is denoted by ${\rm tr} M =
-2\cos\pi\lambda_\nu$.  In this way $C$, or in the case of the torus
$\beta$ through eq.(\ref{betaC}) becomes function of $x$ and
$\delta_\nu=(1-\lambda_\nu^2)/4$. 

This can be considered as an ``off-shell'' $\beta$.  The on-shell
$\beta$ is obtained by choosing $\delta_\nu$ as to have the single
valuedness of the conformal field $\varphi$ under the circuit $M$. As
an illustration we give in Appendix B the perturbative computation of
such $\delta_\nu$. The non perturbative procedure of \cite{ZZ,piatek}
is to fix the value of $\delta_\nu$ from the saddle point which
develops in the semiclassical limit of the quantum expression where
the integration in $dP$ of the dimension $\Delta=
\frac{Q^2}{4}+P^2$ is present.
In all the procedure we work at $\omega_1$ fixed and the $e_j$ in
eq.(\ref{betaC}) are given as functions of the nome $q$ by the
relations \cite{BatemanII,DLMF}

\begin{eqnarray}
&&e_1 = \frac{\pi^2}{12\omega_1^2}(\theta^4_2(0,q)+2~\theta^4_4(0,q))\nonumber\\
&&e_2 = \frac{\pi^2}{12\omega_1^2}(\theta^4_2(0,q)-\theta^4_4(0,q))\nonumber\\
&&e_3 = -\frac{\pi^2}{12\omega_1^2}(2\theta^4_2(0,q)+\theta^4_4(0,q))~.
\end{eqnarray}
Thus we have simply to replace in (\ref{betaC})
\begin{equation}
C(x,\delta_\nu) = C(0,\delta_\nu)+ x C'(0,\delta_\nu)+\frac{x^2}{2}
C''(0,\delta_\nu)+\dots
\end{equation}
where the $C^{(n)}(0,\delta_\nu)$ are the derivatives of $C$ w.r.t. $x$ and
are given by the sphere procedure. We find
\begin{equation}\label{C0}
C(0) = \delta_\nu-\frac{3}{8}
\end{equation}
\begin{equation}
C'(0) = -\frac{\delta_\nu}{2}-\frac{\delta_s}{8}+\frac{3}{8}
\end{equation}
\begin{equation}
C''(0) =
\frac{12\delta_\nu-48\delta_\nu^2+8\delta_\nu\delta_s+\delta_s^2}
{256 \delta_\nu}
\end{equation}
and higher derivatives are reported in Appendix C.

Then exploiting eq.(\ref{betaC}) and 
\begin{equation}
\eta_1 = -\frac{\pi^2}{12\omega_1}\frac{\theta_1'''(0,q)}{\theta_1'(0,q)}=
-\frac{\pi^2}{12\omega_1}(-1+24 q^2+72 q^4+ 96 q^6 + 168 q^8+\cdots)
\end{equation}
we have 
\begin{eqnarray}\label{beta1expansion}
&&\beta -\frac{\eta_1}{\omega_1}\\
&=& \frac{\pi^2}{\omega_1^2}\bigg(\frac{1}{\delta_s}(\frac{1}{4}-\delta_\nu)
-q^2\frac{\delta_s}{2\delta_\nu}+
q^4\frac{\delta_s(-96\delta_\nu^3+3\delta^2_s-5\delta_\nu\delta_s^2+
24\delta_\nu^2(-1+2
\delta_s))}{8\delta_\nu^3(3+4\delta_\nu)}\nonumber\\
&+&q^6\bigg[-(\delta_s(384\delta_\nu^6 + 6\delta_s^4 -
19\delta_\nu\delta_s^4 
- 56\delta_\nu^3\delta_s^2(-1 +
2\delta_s) - 96\delta_\nu^5(-5 + 8\delta_s)\nonumber\\ 
&+&
3\delta_\nu^2\delta_s^2(-16 + 32\delta_s + 3\delta_s^2) +
48\delta_\nu^4(3 - 12\delta_s + 10\delta_s^2)))/
(16\delta_\nu^5(6 + 11\delta_\nu + 4\delta_\nu^2))\bigg]\bigg)\nonumber\\
&+&O(q^8)
\nonumber
\end{eqnarray}

We can now compare the above results with the semiclassical limit
of the quantum conformal blocks.

The conformal block expansion for the torus 1-point function is given
in general by \cite{hadasz,teschnerintro}
\begin{equation}
\langle \phi_{\alpha,\bar\alpha} \rangle_\tau ={\rm tr}(e^{-\tau_I H+i
  \tau_R P}\phi_{\alpha,\bar\alpha}(1,1))
\end{equation}
where $\phi_{\alpha,\bar\alpha}(1,1)$ is the field on the sphere at
point $(1,1)$.

Using
\begin{equation}
H=2\pi(L_0+\bar L_0)-\frac{\pi c}{6},~~~~P=2\pi(L_0-\bar
L_0),~~~~\tilde q= e^{2\pi i \tau}
\end{equation}
we have
\begin{equation}
\langle \phi_{\alpha,\bar\alpha} \rangle_\tau= (\tilde q \bar{\tilde
q})^{-\frac{c}{24}}{\rm tr}(\tilde q^{L_0} \bar{\tilde q} ^{\bar L_0}
\phi_{\alpha,\bar\alpha}(1,1)) ~.
\end{equation}
For Liouville theory 
we have the continuous spectrum \cite{teschnerrev}
\begin{equation}
\Delta = \frac{Q^2}{4}+P^2,~~~~P\in R_+~,
\end{equation}
and for the central charge we have
\begin{equation}
c= 1+6 Q^2=1+6 (b^{-1}+b)^2~.
\end{equation}

Then
\begin{equation}\label{blockexpansion}
\langle \phi_{\alpha} \rangle_\tau =  
\int_{R_+}dP ~~~{\cal F}^\alpha_\Delta(\tilde q)~~{\cal
  F}^{\alpha}_\Delta(\bar{\tilde q})
~~~C^{\alpha}_{\Delta,\Delta} 
\end{equation}
and
\begin{equation}
{\cal F}^\alpha_\Delta(\tilde q)=
\tilde q^{\Delta-\frac{c}{24}}\sum_{n=0} \tilde q^n
F^{\alpha,n}_{\Delta},~~~~\tilde q = e^{2\pi i\tau}= q^2~.
\end{equation}

The conjecture \cite{ZZ,piatek} is that for $b\rightarrow 0$ and heavy
charges i.e. $\alpha=\eta_s/b$,
$\Delta = \frac{1}{b^2}(\frac{1}{4}+p^2)=\frac{\delta_\nu}{b^2}$
${\cal F}^\alpha_{\Delta}(\tilde q)$ exponentiates
\begin{equation}
{\cal F}^\alpha_{\Delta}(\tilde q)\rightarrow
e^{\frac{1}{b^2}f(\eta_s,p,\tilde q)}
\end{equation}
with
\begin{equation}\label{fexpansion}
f(\eta_s,p,\tilde q) = 
(\delta_\nu-\frac{1}{4})\log \tilde q+\lim_{b\rightarrow 0}(b^2 \log
\sum_{n=0}^\infty F^{\alpha,n}_\Delta\tilde q ^n)=
(\delta_\nu-\frac{1}{4})\log \tilde q+\sum_{n=1}^\infty
f_n(\eta_s,p)\tilde q^n
\end{equation}
while at the same time we have 
\begin{equation}
C^{\alpha}_{\Delta,\Delta}\rightarrow
e^{-\frac{1}{b^2}S^{(3)}(\frac{1}{2}-ip,\eta_s,\frac{1}{2}+ip)}
\end{equation}
with $S^{(3)}$ the classical three point action on the sphere.
Then in the $b\rightarrow 0$ limit the integral 
(\ref{blockexpansion}) can be computed
by the saddle point method \cite{ZZ,piatek} giving
\begin{equation}
e^{-\frac{1}{b^2}S^{cl}(\eta_s|\tau)=
}e^{\frac{1}{b^2}(-S^{(3)}(\frac{1}{2}-ip_*,\eta_s,\frac{1}{2}+ip_*)
+f(\eta_s,p_*,\tilde q)+f(\eta_s,p_*,\bar{\tilde q}))}
\end{equation}
$p_*$ being the saddle point given by
\begin{equation}\label{saddlepoint}
\frac{\partial}{\partial p}
\big(-S^{(3)}(\frac{1}{2}-ip,\eta_s,\frac{1}{2}+ip) + f(\eta_s,p,\tilde q) + 
f(\eta_s,p,\bar{\tilde q}) \big)=0~.
\end{equation}

Then exploiting eq.(\ref{saddlepoint}) we have at the saddle point
\begin{equation}\label{qderivative}
\frac{\partial S^{cl}}{\partial \tilde q} = 
-\frac{\partial f}{\partial \tilde q}~.
\end{equation}

Using the classical Ward identity (\ref{classWard}) 
we can then compare expansion (\ref{beta1expansion}) with the expansion of
eq.(\ref{qderivative}). 
From eq.(\ref{fexpansion}) we have
\begin{equation}
f_1 = \lim_{b\rightarrow 0} b^2 F^{1},~~~~f_2 = \lim_{b\rightarrow 0} b^2 \big(F^{2}
-\frac{1}{2} (F^{1})^2\big)
\end{equation}
\begin{equation}
f_3 = \lim_{b\rightarrow 0} b^2 \big(F^{3} -F^{1} F^{2}
+\frac{1}{3}(F^{1})^3\big)~~~{\rm etc}~.
\end{equation}
We verified the agreement of this expansion for the order $q^2$ and
$q^4$ with the classical limit of the quantum conformal blocks given
in \cite{piatek} and as for the $q^6$ term we agree with the results
of \cite{alkalaevbelavin}. With the method described above one can
easily go to higher orders.

\bigskip

\section{Conclusions}

In the present paper we extended the technique to compute the classical
conformal blocks developed in \cite{CCblocks} to the torus topology.
We derived the classical Ward identity connecting the change of
the action under a change of the modulus with the accessory
parameter of the problem, directly from the action.

The classical limit for the quantum action allows to relate the
quantum conformal blocks with the expansion obtained from the
classical Ward identity. The comparison up to the sixth order in the
nome $q$ is performed with success and the calculation can be easily
extended to higher orders.

\bigskip

\section*{Appendix A}

In this appendix we give a short derivation of the identities
eq.(\ref{polyakov}), eq.(\ref{polyakovej}) which where the main
ingredients in proving the classical Ward identity
(\ref{classWard}). The result holds for all hyperelliptic surfaces
with $n$ sources both elliptic and parabolic \cite{HG}.
Let $u_K$ denote the position of the sources and $u_l$ the position of the
the  kinematical singularities which define the hyperelliptic surface
and generalize the $e_l$ of the torus.

In \cite{HG} we used the decomposition of the field $\varphi$
\begin{equation}
\varphi = \varphi_M+\Omega
\end{equation}
where $\Omega$ is a real field which is equal to $-2\eta_K \log(u-u_K)
(\bar u-\bar u_K)$  in finite non overlapping disks
around the sources 
and to $-\frac{1}{2} \log(u-u_l) (\bar u-\bar u_l)$ around the
kinematical singularities,
and equal to $-\frac{3}{2}\log u\bar u$ outside a disk of radius
$R$ which includes all singularities. 
Elsewhere $\Omega$ is defined as a smooth field
which connects smoothly with the field in the described
regions. Notice that $\Omega$ depends on the $u_K, u_l$. Substituting
such decomposition into eq.(\ref{SuTorus}) we obtain \cite{HG}
\begin{equation}\label{Sufinite}
S_u =\frac{1}{2\pi}\lim_{\varepsilon\rightarrow 0}
\int(\frac{1}{2}\partial\varphi_M\wedge\bar
\partial\varphi_M - \varphi_M \partial\bar\partial \Omega-
\frac{1}{2} \Omega \partial\bar\partial \Omega+e^\varphi
du\wedge\bar d\bar u)\frac{i}{2}~.
\end{equation}
Actually the integral is finite in the limit $\varepsilon\rightarrow
0$ and so the limit symbol may be removed but then one has to remember
that $\partial\bar\partial \Omega$ is identically zero in the above
described disks and not e.g. in $D_K$ equal to 
$-2\eta_K\partial_u (1/(\bar u-\bar u_K)) = -2\eta_K\pi\delta^2(u-u_K)$.
We are interested in the derivative of eq.(\ref{Sufinite}) w.r.t. $u_j$.
In was proven in \cite{torusIII,HG} that for the torus with one source
and also for the four point function on the sphere, 
the $\beta$ and the parameter $\kappa$ which appear in the solution 
$\varphi_M$ are real-analytic functions of
the $u_j$ almost everywhere. 
A similar but weaker result holds for higher genus and higher point 
functions \cite{CMSpl,CMS}.

This allows to take in eq.(\ref{Sufinite}) the
derivative operation under the integral sign;
using the equations of motion for $\varphi_M$ and simple integrations
by parts one obtains \cite{HG}
\begin{equation}\label{Szelderivative}
\frac{\partial S_u}{\partial u_l} =
-\frac{i}{4\pi}\oint_{u_l}\Omega_l\partial\varphi_M
+\frac{i}{8\pi}\int
\partial(\varphi_{Ml}\bar\partial\varphi_M)-
\bar\partial(\varphi_{Ml}\partial\varphi_M)
\end{equation}
where the subscript $l$ means the derivative w.r.t. $u_l$.
The local uniformizing variable around $u_l$ is
$s$ with $s^2=u-u_l$. For $Q_s$ we have
\begin{equation}
Q_s= 2 B_l +O(s)~.
\end{equation}
The two independent solution of $f''+Q_s f=0$ 
around $s=0$ are given by
\begin{equation}
f_1= 1+a_1s- B_l s^2+ a_3 s^3+\dots~~~,~~~~~f_2= s+b_2s^2+ b_3 s^3+\dots
\end{equation}
For the $\varphi$ we have
\begin{eqnarray}
\varphi &=& -\frac{1}{2}\log(u-u_l)(\bar u-\bar u_l)-
2 \big[a_1s+\bar a_1\bar s - (B_l+\frac{a_1^2}{2})s^2-(\bar B_l+\frac{\bar
    a_1^2}{2}) \bar s^2
\nonumber\\
&-&\kappa^4 s\bar s + O(s^3)]+{\rm const}~ =
-\frac{1}{2}\log(u-u_l)(\bar u-\bar u_l)+\varphi_M~.
\end{eqnarray}
Thus the contribution of the first integral in eq.(\ref{Szelderivative})
is 
\begin{eqnarray}\label{contributionmoduli}
&&-\frac{i}{4\pi}\oint^d_{u_l}\Omega_l\partial\varphi_M=
\frac{i}{4\pi}\oint \frac{1}{s^2}\bigg(a_1ds-(2B_l+a_1^2) s ds 
-\kappa^4 \bar s d s + O(s^2)ds\bigg)\nonumber\\
&=& - B_l -\frac{a_1^2}{2} = - B_l -\frac{1}{8}(\partial_s\varphi_M)^2 ~.
\end{eqnarray}
Taking into account that around $u_l$,
$\varphi_M=\varphi_M(s|\{u_j\})$ with $s^2=u-u_l$ the second integral
in  eq.(\ref{Szelderivative}) becomes
\begin{equation}\label{boundarycontribution}
-\frac{i}{16}\oint\frac{1}{s}(\partial_s\varphi_M)^2 ds
+\frac{i}{16}\oint
\frac{1}{s}(\partial_s\varphi_M)(\partial_{\bar s} \varphi_M)
d\bar s = \frac{1}{8}(\partial_s\varphi_M)^2+0
\end{equation}
thus leaving the result
\begin{equation}\label{pureBl}
\frac{\partial S_u}{\partial u_l} = - B_l~.
\end{equation}
The proof of eq.(\ref{polyakov}) goes along the same line but is
simpler due to the fact that around $u_K$, $\varphi_M$ is a single
valued function of $u$ and thus the additional term in
eq.(\ref{contributionmoduli}) and the boundary contribution
eq.(\ref{boundarycontribution}) are absent.  For the torus with a
single source placed at $z_s=0$ due to symmetry under reflection
$\phi(z,\bar z)=\phi(-z,-\bar z)$ we have
$\partial_s\varphi_M(e_j)=0$. Not so if the source is at $z_s\neq 0$
with the $u$ defined by eq.(\ref{wpdefinition}) or when we have more
than one source but due to the cancellations discussed above the
result is always (\ref{pureBl}).

\bigskip
\section*{Appendix B}

In this appendix as an illustration we perform the perturbative
computation of the monodromy matrix along the loop $M$ embracing the
cut from $e_3$ to $e_2$. This relates the value of $\delta_\nu$ with
the perturbative value of the accessory parameter $\beta$. As such
$\beta$ is already known we know, perturbatively, the on-shell value
of $\delta_\nu$.

Two unperturbed, i.e. $\delta_s=0$, solutions of the auxiliary equation
\begin{equation}
f''(u)+ Q_u(u) f(u)=0
\end{equation}
are given by \cite{torusI}
\begin{eqnarray}\label{Piu}
&&y_1(u) = [(u-e_1)(u-e_2)(u-e_3)]^\frac{1}{4}\equiv \Pi(u)\nonumber\\
&&y_2(u) = \Pi(u) Z(u)
\end{eqnarray}
with
\begin{equation}\label{Zu}
Z(u) = \int_{e_3}^u\frac{dx}{\sqrt{4(x-e_1)(x-e_2)(x-e_3)}}
\end{equation}
The cuts in eq.(\ref{Piu}) and eq.(\ref{Zu}) are chosen 
from $e_1$ to $+\infty$ and from $e_2$ to $e_3$.
If we go around the cut joining $e_2$ to $e_3$ the monodromies of the 
$y_j(u)$ are
\begin{eqnarray}
&&\tilde y_1(u)= -y_1(u)\\
&&\tilde y_2(u)= \tilde y_1(u) \tilde Z(u) = -y_1(u)(Z(u)+2(\omega_2-\omega_3))=
2(\omega_3-\omega_2) y_1(u)- y_2(u)\nonumber
\end{eqnarray}
with ${\rm tr}M=-2$. To first order perturbation we have 
\begin{eqnarray}
&&\delta y_1(u) = \frac{\delta_s}{4w_{12}}
\bigg(\Pi(u)\int_{e_3}^u\frac{(x+\beta)Z(x)}{\sqrt{(x-e_1)(x-e_2)(x-e_3)}}
dx\nonumber\\
&-&\Pi(u) Z(u) \int_{e_3}^u\frac{x+\beta}{\sqrt{(x-e_1)(x-e_2)(x-e_3)}}dx\bigg)
\end{eqnarray}
and
\begin{eqnarray}
&&\delta y_2(u) = \frac{\delta_s}{4w_{12}}
\bigg(\Pi(u)\int_{e_3}^u\frac{(x+\beta)Z^2(x)}
{\sqrt{(x-e_1)(x-e_2)(x-e_3)}}
dx\nonumber\\
&-&\Pi(u) Z(u)
\int_{e_3}^u\frac{(x+\beta)Z(x)}{\sqrt{(x-e_1)(x-e_2)(x-e_3)}}
dx\bigg)
\end{eqnarray}
where $w_{12}$ is the Wronskian $w_{12}=y_1y_2'-y_1'y_2=1/2$.
The continuation along the circuit $M$ gives
near $e_3$
\begin{equation}
\delta\tilde y_1(u)=\frac{\delta_s}
{4w_{12}}\big((\Omega I_{11}-I_{12})\Pi(u) +I_{11}\Pi(u)Z(u)\big) 
+O((u-e_3)^{\frac{5}{4}})
\end{equation}
and
\begin{equation}
\delta\tilde y_2(u)=\frac{\delta_s}{4w_{12}} 
\big((\Omega I_{12}-I_{22})\Pi(u)+I_{12}\Pi(u)Z(u)\big)+O((u-e_3)^{\frac{5}{4}})
\end{equation}
where
\begin{eqnarray}\label{I11}
&&I_{11} = \oint
\frac{(x+\beta)dx}{\sqrt{(x-e_1)(x-e_2)(x-e_3)}}\nonumber\\
&=&4\int_{e_2}^{e_3}\frac{(x+\beta)dx}{\wp'(z)}=
4\int_{\omega_2}^{\omega_3} (\wp(z)+\beta)dz=
2\Omega\beta+4(\zeta(\omega_3)-\zeta(\omega_2))~,
\end{eqnarray}
with $\omega_3=\omega_1+\omega_2$ and the contour integral is along 
the loop $M$ embracing $e_3,e_2$.
$\zeta(z)$ is the elliptic zeta-function and $\Omega=
2(\omega_2-\omega_3)$. The values of 
$I_{12}$ and $I_{22}$ do
not intervene in the following calculation. 

Thus we have for the trace of $M$ to first order
\begin{equation}
{\rm tr}M=-2 \cos(\pi \lambda_\nu)= -2+\frac{\delta_s}{4w_{12}} \Omega I_{11}~.
\end{equation}
Expanding we have to first order in $\delta_s$
\begin{equation}\label{firstorderdeltanu}
\delta_\nu =\frac{1-\lambda_\nu^2}{4} =
\frac{1}{4}(1-\frac{\delta_s}{4 w_{12}\pi^2} \Omega I_{11})
\end{equation}
where $I_{11}$ is given by eq.(\ref{I11}) and gives the relation 
between the accessory parameter $\beta$ and the trace of the
monodromy $M$.
The direct perturbative computation of $\beta$ \cite{torusI} gives
\begin{equation}
\beta = \frac{\bar\omega_2\zeta(\omega_1)-\bar\omega_1\zeta(\omega_2)}
{\bar\omega_2\omega_1-\bar\omega_1\omega_2}
=\frac{\eta_1}{\omega_1}-\frac{\pi}{4\omega_1^2\tau_I} ~.
\end{equation}

Substituting in (\ref{firstorderdeltanu}) it gives
\begin{equation}
\delta_\nu =
\frac{1-\lambda_\nu^2}{4}=\frac{1}{4}\big(1+\frac{\delta_s}{\pi\tau_I}\big)
> \frac{1}{4}
\end{equation}
showing that the monodromy $M$ is hyperbolic.

\section*{Appendix C}

We report here the values of the derivatives of the accessory
parameter $C(x,\delta_\nu)$ which were used in the text to compute the
torus conformal blocks
\begin{equation}
C^{(3)}(0,\delta_\nu)
=3(-48\delta_\nu^2+\delta_s^2+4\delta_\nu(3+2\delta_s))/(512\delta_\nu)
\end{equation}
\begin{eqnarray}
&& C^{(4)}(0,\delta_\nu)
=\big(-9~\delta_s^4/(65536~\delta_\nu^3) +
15~\delta^4_s/(65536~\delta^2_\nu)\nonumber\\
&-&9~\delta^2_s(-41+ 2 ~\delta_s)/(8192~\delta_\nu)
+3~(729+492 ~\delta_s+86 ~\delta_s^2)/4096\nonumber\\
&+&3\delta_\nu(-243+82~\delta_s)/512-729~\delta_\nu^2/256\big)
/(3+4 ~\delta_\nu)
\end{eqnarray}
\begin{eqnarray}
&&C^{(5)}(0,\delta_\nu)\nonumber\\
&=&(-15(88320 ~\delta_\nu^5 + 9 ~\delta_s^4 - 15 ~\delta_\nu ~\delta_s^4 + 
   24 ~\delta_\nu^2 ~\delta_s^2(-59 + 6 ~\delta_s) \nonumber\\ 
&-&128 ~\delta_\nu^4(-345 + 118 ~\delta_s) - 16 ~\delta_\nu^3
    (1035 + 708 ~\delta_s + 130 ~\delta_s^2)))/\nonumber\\
&&(131072 ~\delta_\nu^3(3 + 4 ~\delta_\nu))
\end{eqnarray}
\begin{eqnarray}
&&C^{(6)}(0,\delta_\nu)=\nonumber\\
&&(-45(17190912~\delta_\nu^8 - 6 ~\delta_s^6 + 19 ~\delta_\nu ~\delta_s^6 + 
   56 ~\delta_\nu^3 ~\delta_s^4(-111 + 2 ~\delta_s) -\nonumber\\ 
&&   2048 ~\delta_\nu^7(-20985 + 1454 ~\delta_s) - 3 ~\delta_\nu^2 ~\delta_s^4
    (-1776 + 32 ~\delta_s + 3 ~\delta_s^2) - \nonumber\\
&& 16 ~\delta_\nu^4 ~\delta_s^2
    (34893 - 5316 ~\delta_s + 305 ~\delta_s^2) - 
   256 ~\delta_\nu^6(-54561 + 31988 ~\delta_s + 1675 ~\delta_s^2) +\nonumber\\ 
&&   256 ~\delta_\nu^5(-25182 - 17448 ~\delta_s - 4439 ~\delta_s^2 + 
168 ~\delta_s^3)))/\nonumber\\
&& (16777216 ~\delta_\nu^5(6 + 11 ~\delta_\nu + 4 ~\delta_\nu^2))
\end{eqnarray}

\eject

\vfill

%180626

\end{document}